\documentclass[preprint,onecolumn,showpacs,preprintnumbers]{revtex4}
\usepackage{graphicx}
\usepackage{dcolumn}
\usepackage{bm}
\usepackage{amsmath}
\usepackage{amsfonts}

\newcommand{\theq}{\theta_{\text{eq}}}
\newcommand{\vct}[1]{\mathbf{#1}}

\begin{document}

\title{Motion of nanodroplets near edges and wedges}

\author{A.  Moosavi} 
\affiliation{Max-Planck-Institut f\"ur Metallforschung, Heisenbergstr. 3,
D-70569 Stuttgart, Germany,} 
\affiliation{Institut f\"ur Theoretische und Angewandte Physik,
Universit\"at Stuttgart, Pfaffenwaldring 57, D-70569
Stuttgart, Germany}
\author{M. Rauscher}
\email{rauscher@mf.mpg.de}
\affiliation{Max-Planck-Institut f\"ur Metallforschung,
Heisenbergstr. 3, D-70569 Stuttgart, Germany,}
\affiliation{Institut f\"ur Theoretische und Angewandte Physik,
Universit\"at Stuttgart, Pfaffenwaldring 57, D-70569
Stuttgart, Germany}
\author{S. Dietrich}
\affiliation{Max-Planck-Institut f\"ur Metallforschung,
Heisenbergstr. 3, D-70569 Stuttgart, Germany,}
\affiliation{Institut f\"ur Theoretische und Angewandte Physik,
Universit\"at Stuttgart, Pfaffenwaldring 57, D-70569
Stuttgart, Germany}

\date{\today}

\begin{abstract}

Nanodroplets residing near wedges or edges of solid substrates exhibit a
disjoining pressure induced dynamics. Our nanoscale hydrodynamic 
calculations reveal that non-volatile droplets are attracted or repelled 
from edges or wedges depending on details of the corresponding 
laterally varying disjoining pressure generated, e.g., 
by a possible surface coating.
\end{abstract}

\maketitle

Wetting phenomena \cite{deGennes:1985,Dietrich:1988} and thin film dynamics
\cite{Oron:1997} on rough \cite{Netz:1997} or topographically structured
\cite{Gang:2005,Bruschi:2002} substrates have been studied in detail both 
experimentally \cite{Gang:2005,Bruschi:2002,Klier:2005,Ondarcuhu:2005} 
and theoretically \cite{Netz:1997,Rejmer:1999,Rascon:2000}\/. 
To a large extent these investigations
are motivated by the fact that generically substrates are rough, in
particular those playing an important role in technological processes
such as oil recovery, coating, lubrication, and paper processing. A
second driving force for the progressive application of these
micro- and nanofluidic processes stems from the lab-on-a-chip concept
which integrates pipes, pumps, reactors, and analyzers into a single
device allowing for a cost efficient handling of minute amounts of
liquid containing, e.g., DNA or proteins \cite{Karniadakis:2005}\/.

Most theoretical investigations of wetting of structured substrates
have been concerned with thermal equilibrium. Only recently efforts
have been  made towards understanding the corresponding dynamics.
Using various numerical techniques, time-dependent free surface flow
over topographic features has been investigated \cite{Gaskell:2006,
Gramlich:2004, Gaskell:2004, Bielarz:2003}\/. However, the
applicability of these results to the nanoscale is impeded because
the effect of intermolecular interactions, relevant at the nanoscale,
either has not been considered \cite{Gaskell:2006, Gramlich:2004} or
only in a rather crude way \cite{Gaskell:2004, Bielarz:2003}\/. 

In contrast, here we study nanoscale fluid
dynamics on topographical surface structures by properly taking into account
the spatial variation of the long-ranged intermolecular interactions.
We focus on edges and wedges as two paradigmatic
geometric structures and provide a detailed dynamical study of their
effects on nanodroplets positioned in their vicinity. We consider a
partially wetting, non-volatile, and incompressible liquid film
composed of a nanodroplet on top of a precursor wetting layer on a
solid substrate.  We take into account that the surface of the edge 
is possibly covered with a thin solid coating
layer of a material different from the bulk 
substrate which allows one to tune the substrate
potential \cite{Bauer:1999}.
Our analysis shows that the
dynamics of droplets on such structures depends on subtle details of
the substrate properties. 

\begin{figure}
\includegraphics[width=0.49\linewidth]{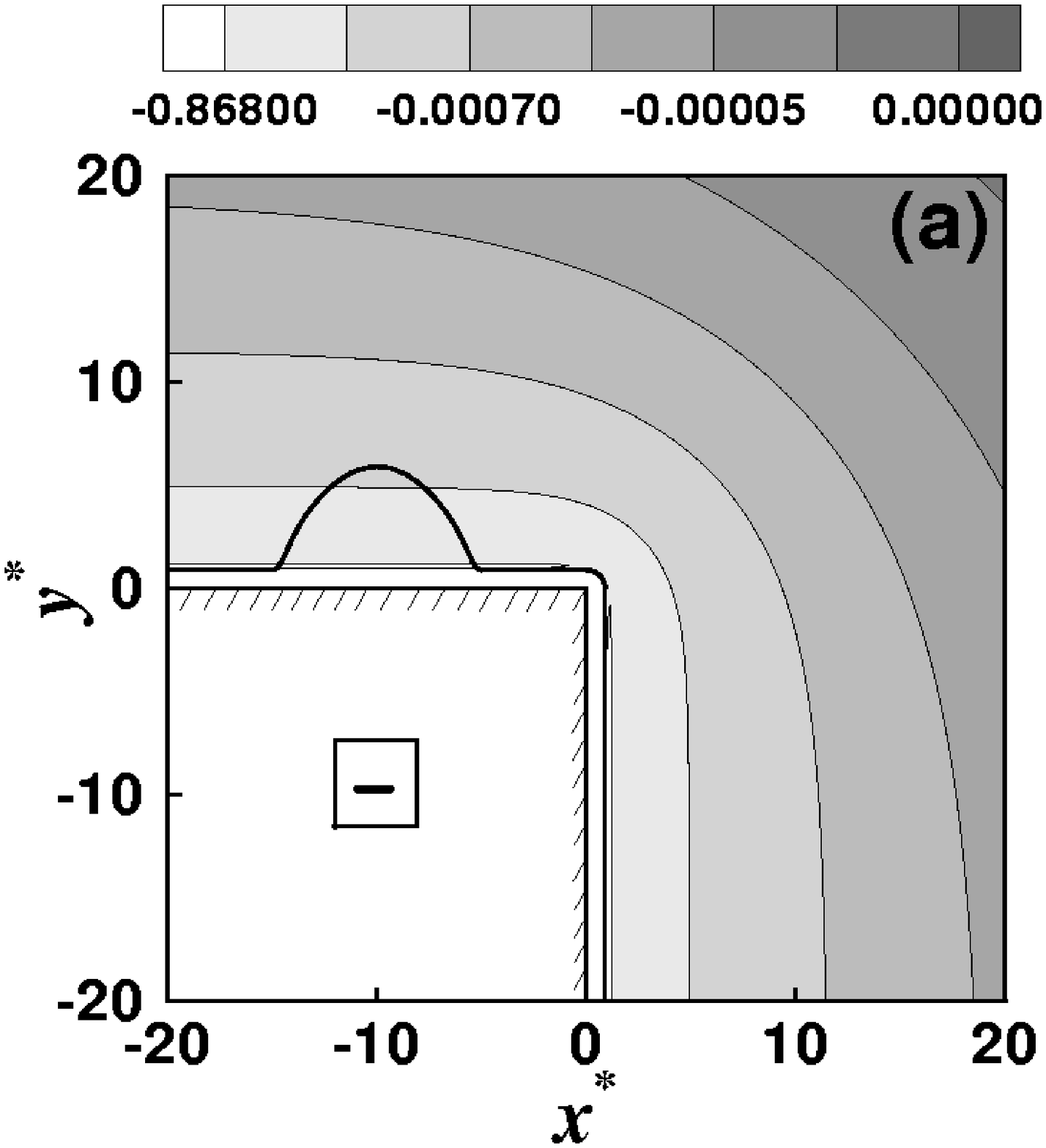}
\includegraphics[width=0.49\linewidth]{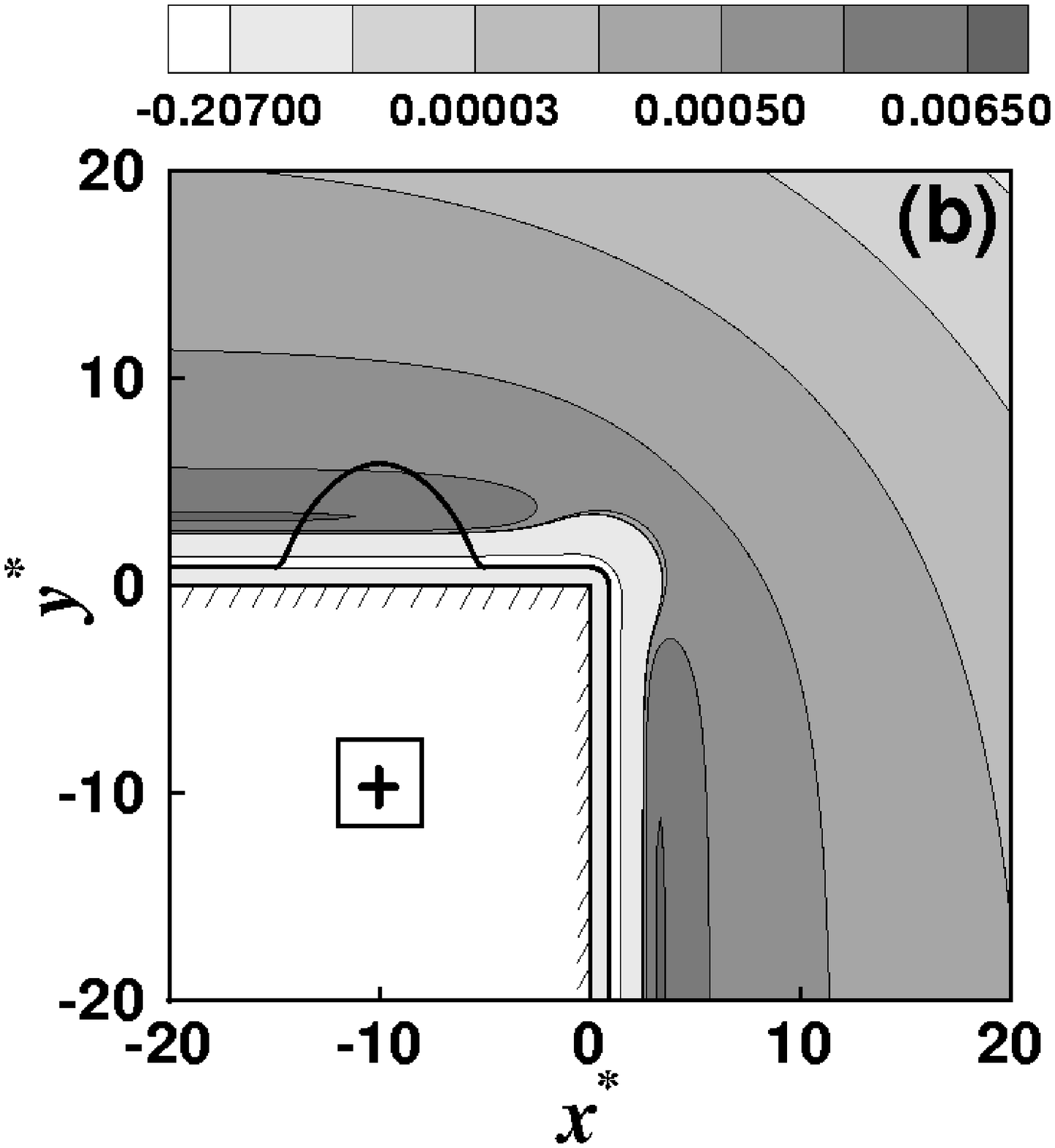}
\caption{\label{fig1} Edge configuration. 
A nanodroplet (full line) 
is placed near an edge and thus exposed to the action  of the laterally 
varying DJP. Contour plots of the corresponding dimensionless DJP for 
the minus ((a), $B=-1$, $C=1$) and plus ((b), $B=-2.5$, $C=1$) 
case are shown (see Eq.~(\protect\ref{djpfar})). 
The edge surfaces are at $x^*=0$ and $y^*=0$, respectively.} 
\end{figure}

\begin{figure}[b]
\includegraphics[width=0.49\linewidth]{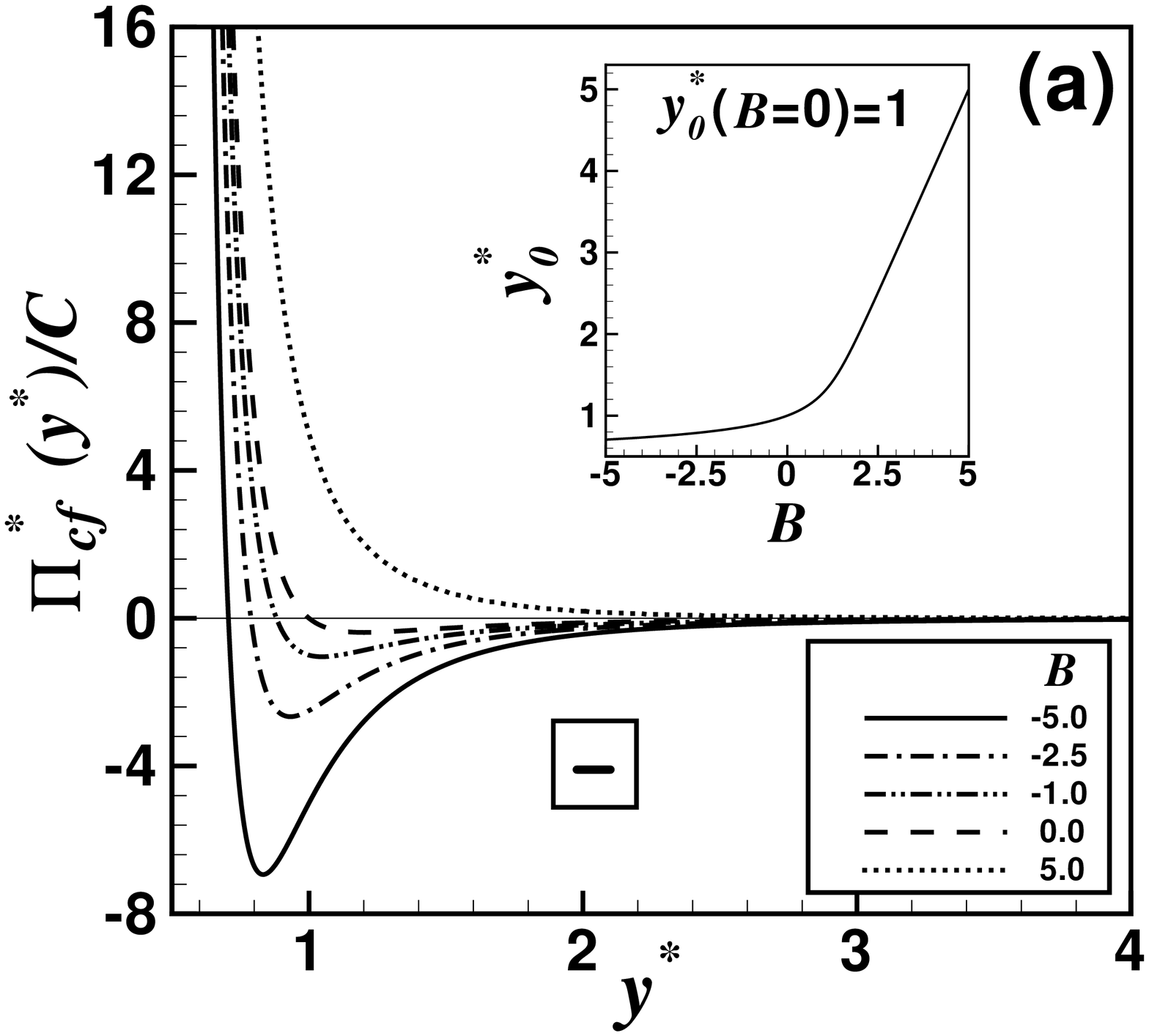}
\includegraphics[width=0.49\linewidth]{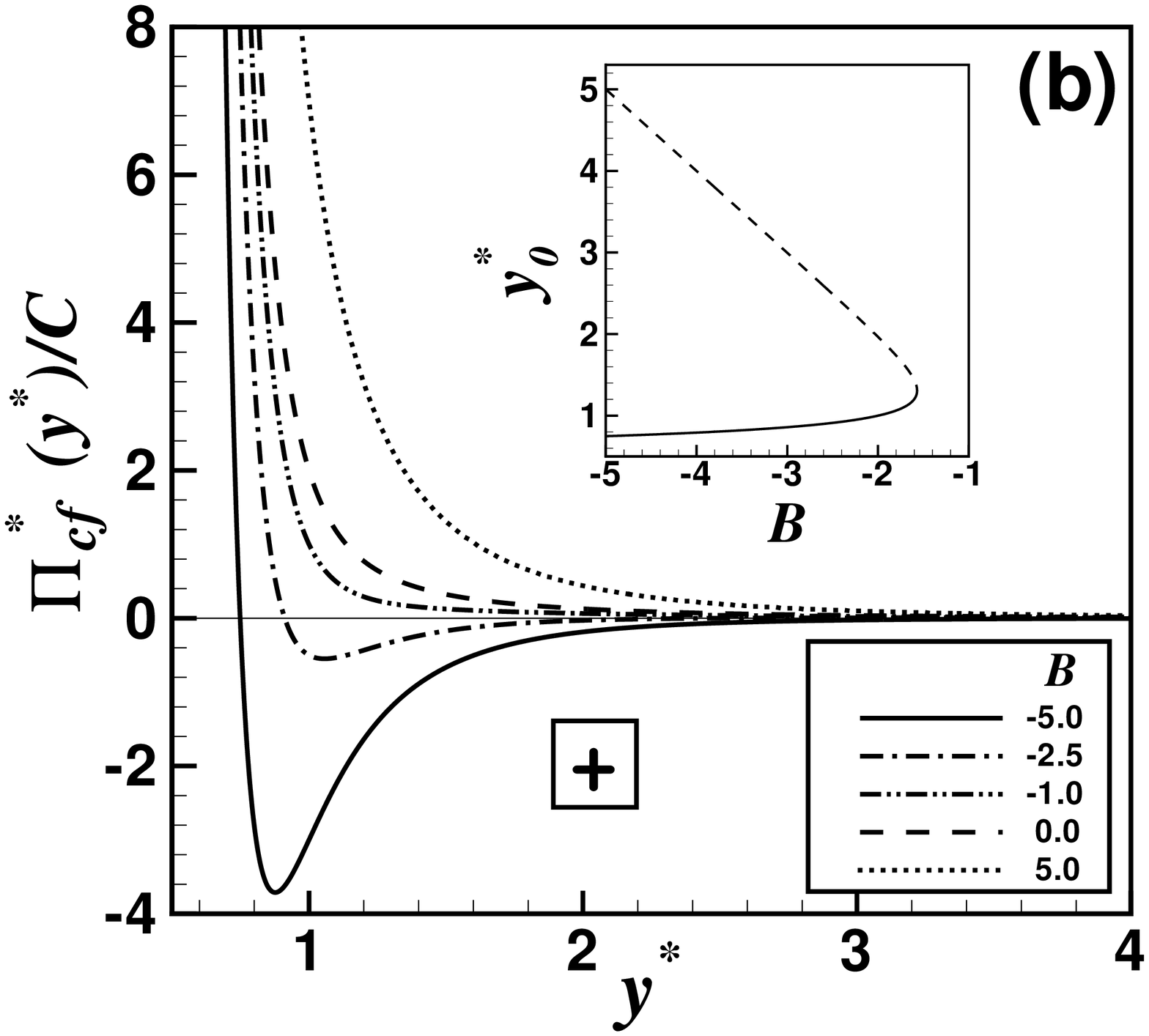}
\caption{\label{fig2} 
Rescaled DJP far from the edge 
in the plus (a) and the minus (b) case for different $B$ (see
Eq.~(\protect\ref{djpfar}))\/. The insets show the positions of the
zeros of $\Pi_{cf}^*$ as a function of $B$\/.
In the
plus case $\Pi_{cf}^*$ has zeros only for $B<-1.57$\/. 
The full curves in the insets correspond
to the equilibrium wetting layer thickness.
} 
\end{figure}

For given intermolecular interactions, 
based on density functional theory the free energy of a 
prescribed liquid film configuration in contact
with, e.g., a wedge-like substrate is a functional of the
liquid-vapor interface shape \cite{Napiorkovski:1992}\/ 
which can be expressed in terms of the 
so-called effective interface
potential $\omega$ \cite{Napiorkovski:1992,Dietrich:1991}. 
For a flat substrate and film thickness $h$ one 
has  $\omega(h) = -\int_h^\infty \Pi(y)\,dy$ where $\Pi$ is 
the disjoining pressure (DJP). Since the equation of motion can be expressed 
in terms of $\Pi$, we determine $\Pi$ directly following Ref. \cite{Bauer:1999}\/.
Assuming Lennard-Jones type pair potentials $V_{\alpha
\beta}(r)={M_{\alpha \beta}}/{r^{12}}-{N_{\alpha
\beta}}/{r^6}$, where $M_{\alpha \beta}$
and $N_{\alpha \beta}$ are material parameters, 
and $\alpha$ and $\beta$ 
relate to liquid ($l$), substrate ($s$), or coating ($c$) particles,
the DJP corresponding to chemically homogeneous substrate is given by \cite{Robbins:1991}
\begin{equation}
\label{eq1}
\Pi({\bf {r}})=
\int_{\Omega_s}
{\left[\rho_l^2\,V_{ll}({\bf{r}}-{\bf{r'}})
-\rho_l\,\rho_s\,V_{sl}({\bf{r}}-{\bf{r'}})\right]}
\,d^3r'
\end{equation} 
with ${\bf{r}}=(x,y,z)\in \mathbb{R}^3$\/ and $\rho_l$ and $\rho_s$ are the
number densities of the liquid and the substrate, respectively.
$\Omega_s$ is the substrate volume. Equation~(\ref{eq2}) assumes that the vapor
density is so low that its contribution to $\Pi$ can be neglected.
For a non-coated $e$dge $\Omega_s=\lbrace {\bf{r}} \in \mathbb{R}^3
\mid x,y\leq 0 , z\in \mathbb{R} \rbrace $ 
(see Fig.~\ref{fig1}) this yields 

\begin{equation}
\label{eq2}
\Pi_e(x,y)= \int_{\Omega_s} \frac{\Delta M}{\left|
\vct{r}-\vct{r}'\right|^{12}}\,d^3r' -
\int_{\Omega_s} 
\frac{\Delta N}{\left|
\vct{r}-\vct{r}'\right|^6 }\,d^3r'
\end{equation} 
where $\Delta M= \rho_l^2 M_{ll}-\rho_l\rho_s M_{sl}$ and $\Delta N=\rho_l^2
N_{ll}-\rho_l\rho_s N_{sl}$\/. The first (second) 
term dominates near (far from) the substrate.
On a flat and homogeneous substrate the 
equilibrium thickness of the wetting layer is
given by the zero of $\Pi$ which corresponds to the minimum of $\omega$\/.
For the interactions considered here, $\Delta M\ge0$ 
is a necessary condition for the occurrence 
of an equilibrium wetting layer of nonzero thickness. 
Both integrals in Eq.~(\ref{eq2}) can be calculated analytically and
one obtains the DJP as the corresponding difference of 
two contributions $\Pi_e=\Pi_e^6-\Pi_e^{12}$\/.
In order to enrich the model we consider in addition the case that the substrate 
is covered by a thin coating layer of thickness $d$\/. 
Actual coating layers have a more complicated
structure, in particular around edges or wedges, which depends on the
specific combination of coating and substrate material as well as
how the coating is deposited. Such details do influence the
motion of droplets very close to the edge but corresponding 
calculations carried out by us 
indicate that a simple model with rather small $d$ captures 
correctly the dynamics at lateral distances 
from the edge larger than $d$\/.
The contribution of a thin coating layer to the DJP can be determined
as above, assuming a van der Waals type interaction between the
coating and the liquid particles. We do not consider an additional
repulsive part of the liquid-coating interaction as this is 
shorter ranged ($\sim y^{-10}$) than the corresponding part
$\Pi_e^{12}\sim y^{-9}$ \cite{Bauer:1999, Dietrich:1991}\/.
The contribution $\Pi_c^u(x,y)$ of the coating layer on the
horizontal $u$pper part of the edge $\{(x,y,z)|x<0,y=0\}$ to the DJP 
can also be calculated analytically. 
By symmetry, the contribution of the vertical coating layer 
is $\Pi^u_c(y,x)$\/. Thus the DJP with the coating is 
\begin{equation}
\label{djp}
\Pi_{ce}(x,y)=\Pi_e(x,y)+\Pi^u_c(x,y)+\Pi^u_c(y,x).
\end{equation} 
Far from the edge, i.e., for $x\rightarrow -\infty$, the DJP
reduces to that of a coated $f$lat substrate: 
$\Pi_{cf}(y)={\pi\Delta M}/{(45y^9)}-{\pi\Delta
N}/{(6y^3)}+{\pi \Delta N' d}/{2y^4}$, with  $\Delta N'=\rho_l^2
N_{ll}-\rho_l\rho_c N_{cl}$\/ measuring the
interaction strength of the coating layer. 
We introduce dimensionless quantities (marked by $*$)  such that lengths
are measured in units of $b={[2\Delta M/(15\,|\Delta N|)]}^{1/6}$ which 
for $\Delta N>0$ is the equilibrium wetting film thickness on the
uncoated flat substrate. For the relation between $b$ and the 
equilibrium wetting film thickness on the coated substrate see the 
insets in Fig. \ref{fig2}. The DJP is measured in units of the 
ratio $\sigma/b$ where $\sigma$ is liquid-vapor surface tension.
Thus the dimensionless DJP $\Pi_{cf}^*=\Pi_{cf}b/\sigma$ far from the edge has the form 
\begin{equation}
\label{djpfar}
\Pi_{cf}^*(y^*)={C}\left(\frac{1}{{y^*}^9} \pm
\frac{1}{{y^*}^3} + \frac{B}{{y^*}^4}\right).
\end{equation} 
The dimensionless amplitude $C=A\,b/\sigma$, where
$A=\pi{(\Delta M/45)}^{-1/2}{(|\Delta N|/6)}^{3/2}$, compares
the strengths of the effective intermolecular forces in the uncoated case 
and of the surface tension forces. The amplitude $B=\pi \Delta
N' d/(2\,A\,b^4)$,
which can be positive or negative, measures the strength of the
coating layer. Note that an analysis of the DJP, 
which is more refined than Eq.~(\ref{eq1}), yields $B\neq0$ even 
in the absence of a coating layer \cite{Dietrich:1988,Dietrich:1991}; 
therefore in the following we consider $B$ as an independent parameter. 
In the second term on the right hand side of Eq.~(\ref{djpfar}) 
the upper (lower) sign corresponds to $\Delta N<0$ $(\Delta N>0)$. 
In the following we shall refer to these cases as the plus and the 
minus case (see Fig. \ref{fig2}).
In the minus case, the DJP has
a single zero for all $B$, while in the plus case 
there is no zero for $B>-1.57$ and two zeros for $B<-1.57$\/. A
DJP with only one zero corresponds to a
substrate which can undergo a second-order wetting transition,
while two zeros indicate a possible first-order wetting transition. 
The absence of a zero corresponds to complete wetting \cite{Dietrich:1988}\/. 
The dimensionless form of the DJP for a coated edge
is given by
\begin{multline}
\Pi_{ce}^*(x^*,y^*)=C\,\bigg\{
\frac{45\,\Pi_e^{12}(x^*,y^*)}{\pi\,\Delta M}
\pm\frac{6\,\Pi_e^{6}(x^*,y^*)}{\pi\,|\Delta N|}\\
+\frac{2\,B\,[\Pi_c^{u}(x^*,y^*)+\Pi_c^{u}(y^*,x^*)]}{\pi\,|\Delta
N'|}\bigg\}.
\end{multline} 
The physically possible ranges of $C$ and $B$ in these cases can be inferred from 
considering the macroscopic equilibrium contact angle $\theq$ given by 
$\cos\theq=1+\omega_{cf}^*(y_0^*)$ 
with $y^*_0$ being the minimum of the corresponding effective interface potential
$\omega_{cf}^*(y^*)=\int_{y^*}^{\infty}\Pi_{cf}^*(y')\,dy'$
\cite{Dietrich:1988}\/.
$|\cos\theq|\le 1$ implies $B<-1.87$ for the plus case so that the 
disjoining pressure has two zeros. In the minus case there is no limitation for $B$.

Figures \ref{fig1}(a) and (b) show typical examples of the
DJP at an edge for both the minus and the plus case, respectively. 
If a droplet is placed near the edge, it is exposed to the lateral 
gradient of the disjoining pressure resulting in a lateral force on the droplet. 

\begin{figure}[t]
\includegraphics[width=0.49\linewidth]{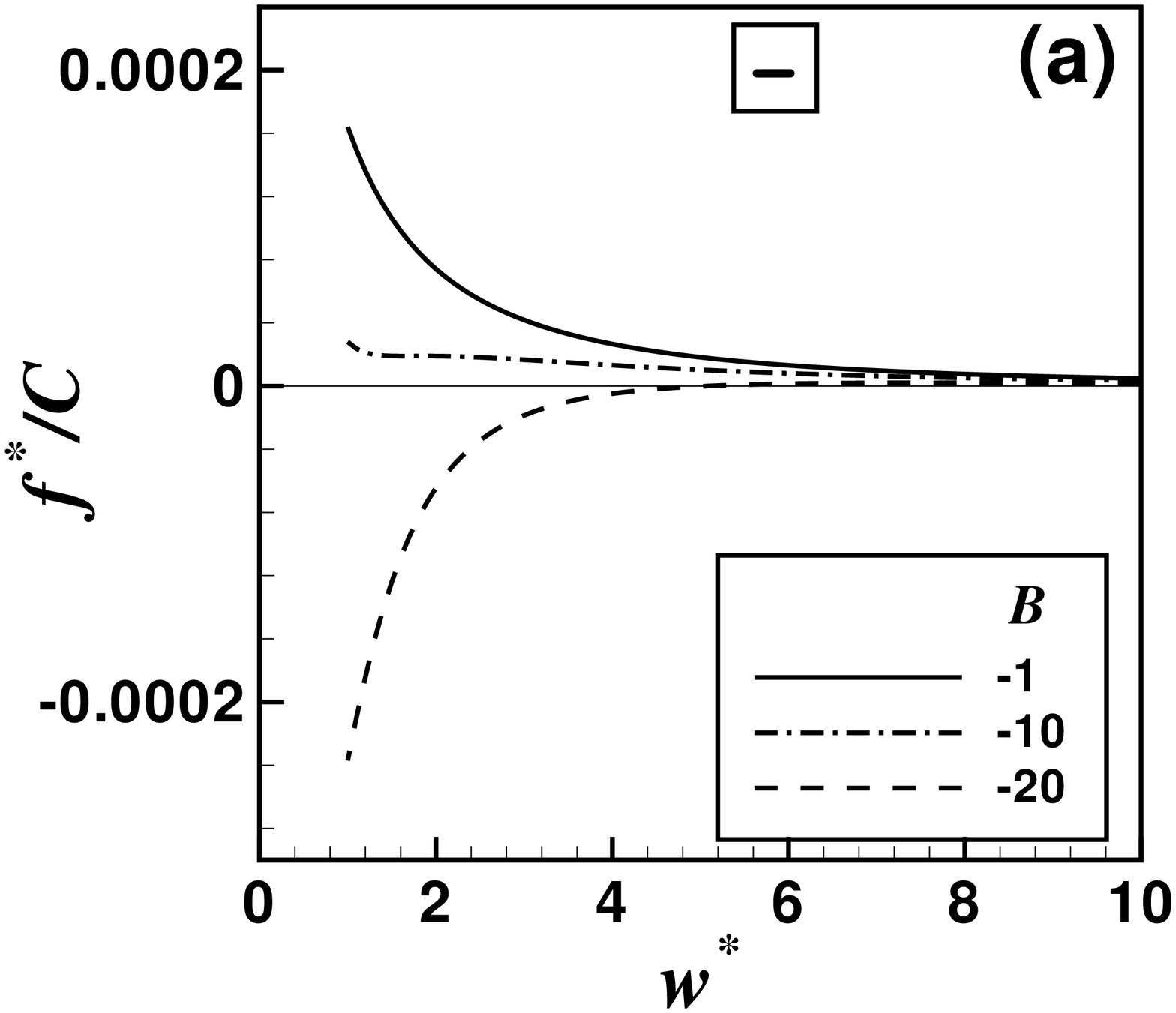}
\includegraphics[width=0.49\linewidth]{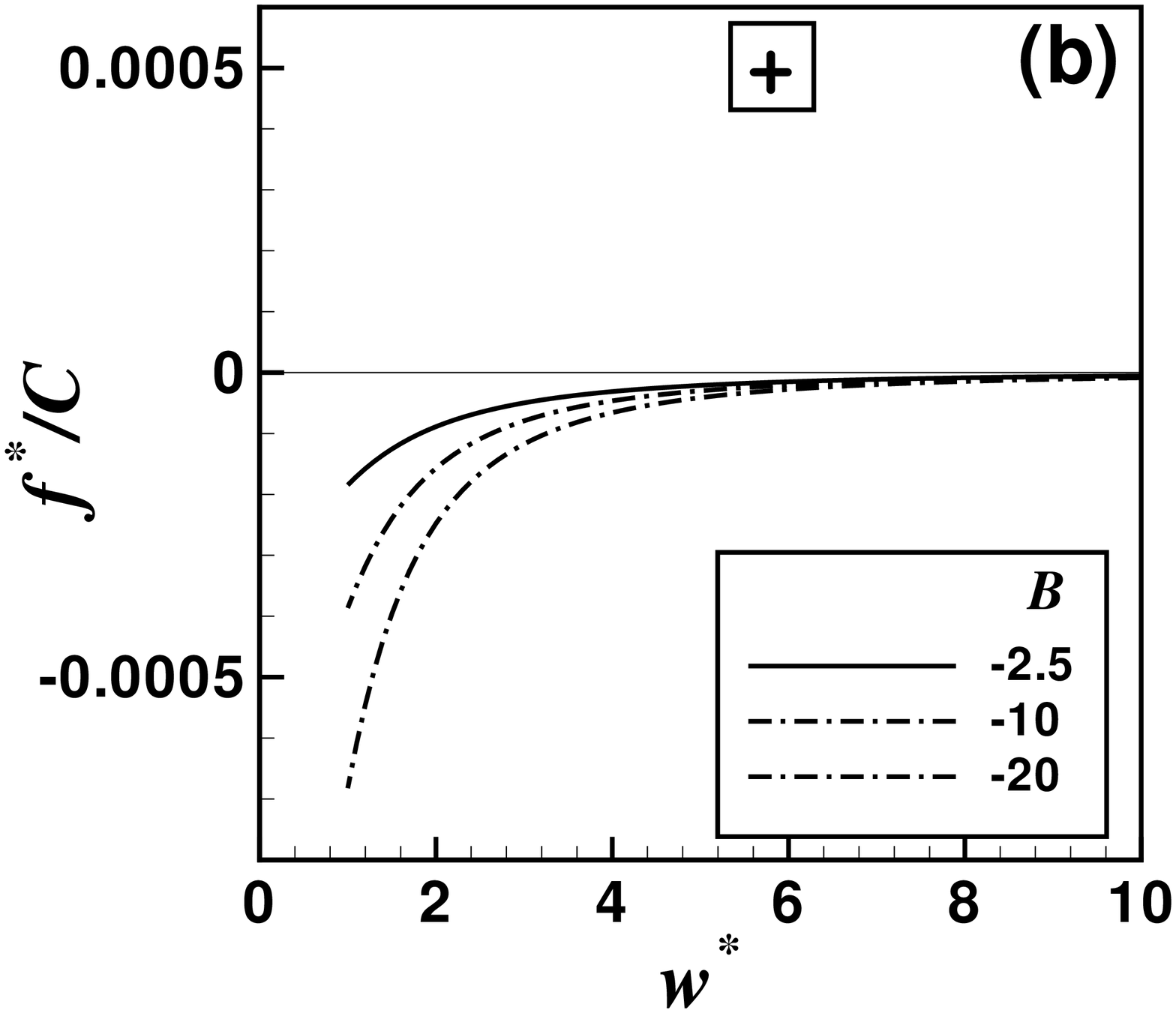}
\caption{\label{fig3} 
Normalized DJP induced lateral force densities $f^*$ acting
on an initial droplet configuration (see Fig.~\ref{fig1}) of
height $a^*=15$ and at distance $w^*$ (see Fig. \ref{fig4}) from an edge 
for the minus (a) and the plus (b) case.
} 
\end{figure}

\begin{figure}[b]
\includegraphics[width=\linewidth]{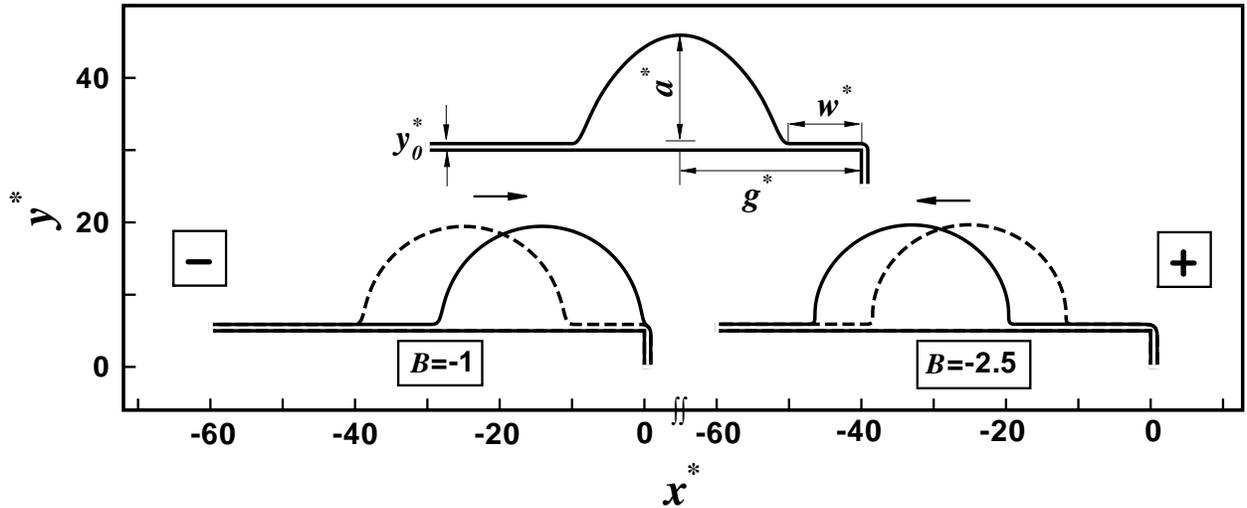}
\caption{\label{fig4} Motion of a droplet ($a^*=15$,
$w^*=10$) near an edge for the minus
(lower left, $B=-1$, $C=1.27$) and the plus (lower right, $B=-2.5$,
$C=4.23$) case. For $y_0^*$ see the insets in Fig. \ref{fig2}. 
The upper figure shows the initial droplet shape.
Shown are interface profiles at $t^*=339$ (dashed) and $t^*=20893$
(solid) for the minus case and at $t^*=210$ (dashed) and $t^*=93699$
(solid) for the plus case. The chosen values of $B$ and $C$ provide
$\theq=90^\circ$\/.}
\end{figure}

In the following we restrict our analysis to 
ridges translationally invariant along $z$, expecting that 
our arguments and conclusions carry over qualitatively to
actual three-dimensional droplets. 
For a ridge, the lateral force density on the droplet is
$f^*=\frac{1}{\Omega^*_d}\,\int_{\partial\Omega^*_d}
\Pi_{ce}^*(x^*,y^*)\,n_{x^*}\,d{S^*}$\/.
$\partial\Omega^*_d$ and $\Omega^*_d$ are the dimensionless droplet
surface and volume, respectively, and $n_{x^*}$ is the $x^*$-component of the unit 
surface normal vector pointing outward, $\mathbf{\hat{n}}$. 
In Fig. \ref{fig3} this force density, 
estimated by the force density on a parabolic
droplet crossing over smoothly to a wetting layer of thickness $y_0^*$
(discussed as initial configuration for the Stokes dynamics below,
see also Fig.~\ref{fig1}), is plotted as a function of the
distance $w^*$ (see Fig. \ref{fig4}) between the three-phase contact line 
and the edge for different values of $B$.
For the minus case there is a critical value $B_c\simeq -10$ which depends
weakly on the drop size.
For $B>B_c$ the force is always positive and
increases towards the edge. Thus one expects the droplet to move towards 
the edge. However, for $B<B_c$ the force changes sign from plus to minus 
upon approaching the
edge and the droplet will stop at a certain distance from the edge which
increases with decreasing $B$\/. 
In the plus case the force is always negative, i.e., pointing away from the
edge, and increases upon approaching the edge. Therefore in this case one
expects a droplet to move away from the edge. 

We analyze the liquid flow of this motion in terms of a two-dimensional 
Stokes equation. In dimensionless form the continuity and Stokes equation
read
\begin{equation}
\bm{\nabla}\cdot\mathbf{u}^*=0\quad\text{,}\quad
\label{eqstokes}
\bm{\nabla}^2{\mathbf{u}^*}=\frac{1}{C}\,\bm{\nabla}\left(p^*+\Pi^*\right),
\end{equation} 
where $\mathbf{u}^*=(u^*_{x^*},u^*_{y^*})$ is the velocity
vector and  $p^*$ is the hydrostatic pressure. 
The velocity and time scales are $A\,b/\mu$ and
$\mu/A$, respectively, with the viscosity $\mu$\/. 
Lengths and pressure have been
scaled with $b$ and $\sigma/b$, respectively. 
At the liquid-vapor interface mass conservation
yields $v^*= \mathbf{\hat{n}}\cdot\mathbf{u}^*$ for the normal 
interface velocity $v^*$\/. 
At the liquid-solid interface a no-slip condition is applied and
there is no flux into the impermeable substrate. Along the
liquid-vapor interface the tangential
stresses are zero (neglecting the viscosity of the vapor phase)
and normal stresses are balanced by the pressure, the DJP, and the
surface tension \cite{Kelmanson:1983} (in dimensional form $\bm{\tau}\cdot \mathbf{\hat{n}} -
p \mathbf{\hat{n}} = \sigma\kappa\mathbf{\hat{n}}$ with the 
stress tensor $\bm{\tau}$ and the mean curvature $\kappa$)\/.
There is no flux through the boundaries of the box used for the numerical
calculations, guaranteeing mass conservation. We solve these equations
numerically with a standard biharmonic boundary integral method
\cite{Kelmanson:1983} for initial droplet shapes of the form 
$y^*(x^*;t=0)= y_0^*+a^*\lbrace 1-[(x^*-g^*)/a^*]^2]
\rbrace^{{\mid x^*-g^*\mid}^m+1}$.
$a^*$ is the droplet height in the center and half the base width, 
and $g^*$ is the distance of the droplet center from the edge. In this study
$m$ was chosen to be $10$\/. At $t=0$ the droplet is positioned with its
three-phase contact line $(x^*=-(g^*-a^*),\,y^*=y_0^*,z^*)$ at a
distance $w^*=g^*-a^*$ away from the edge (see Fig. \ref{fig4}). 

Figure~\ref{fig4} illustrates the difference of the motion of
the droplets in the minus and the plus case for values of $B$ and $C$
such that $\theq=90^\circ$ is maintained. ($B$ and $C$ can be varied such 
that $\theta_{eq}$ remains constant. In this sense the present dynamics is 
complementary to the motion of droplets caused by chemically 
generated contact angle gradients \cite{Raphael:1988}.) The droplet 
moves towards the edge in the minus case and away from the edge in
the plus case. The motion towards the edge in the minus case
accelerates but comes to a stop when the right three-phase contact line of the drop
reaches the edge. In the plus case, the drop motion decelerates but does not stop. 
\begin{figure}
\includegraphics[width=0.49\linewidth]{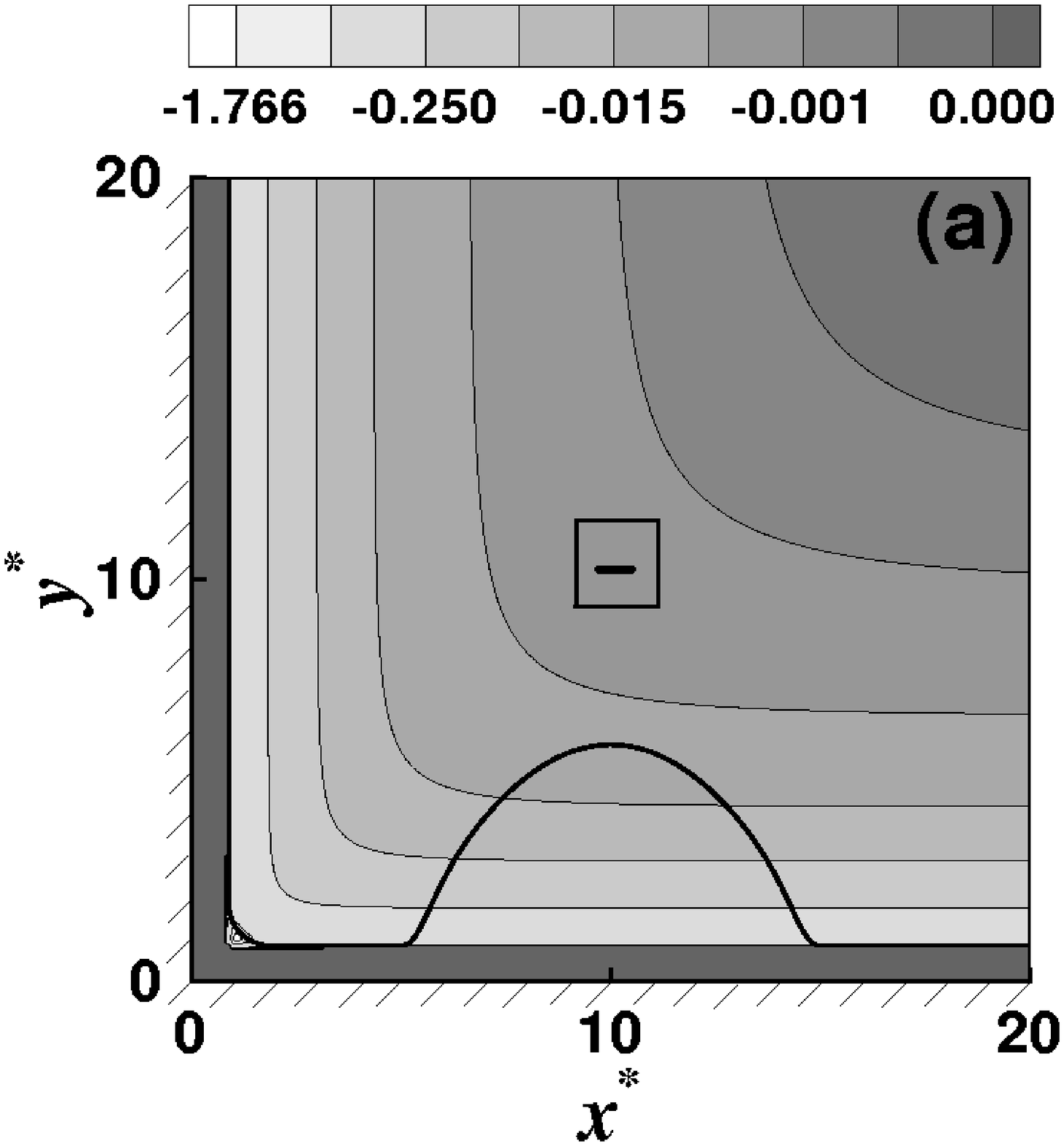}
\includegraphics[width=0.49\linewidth]{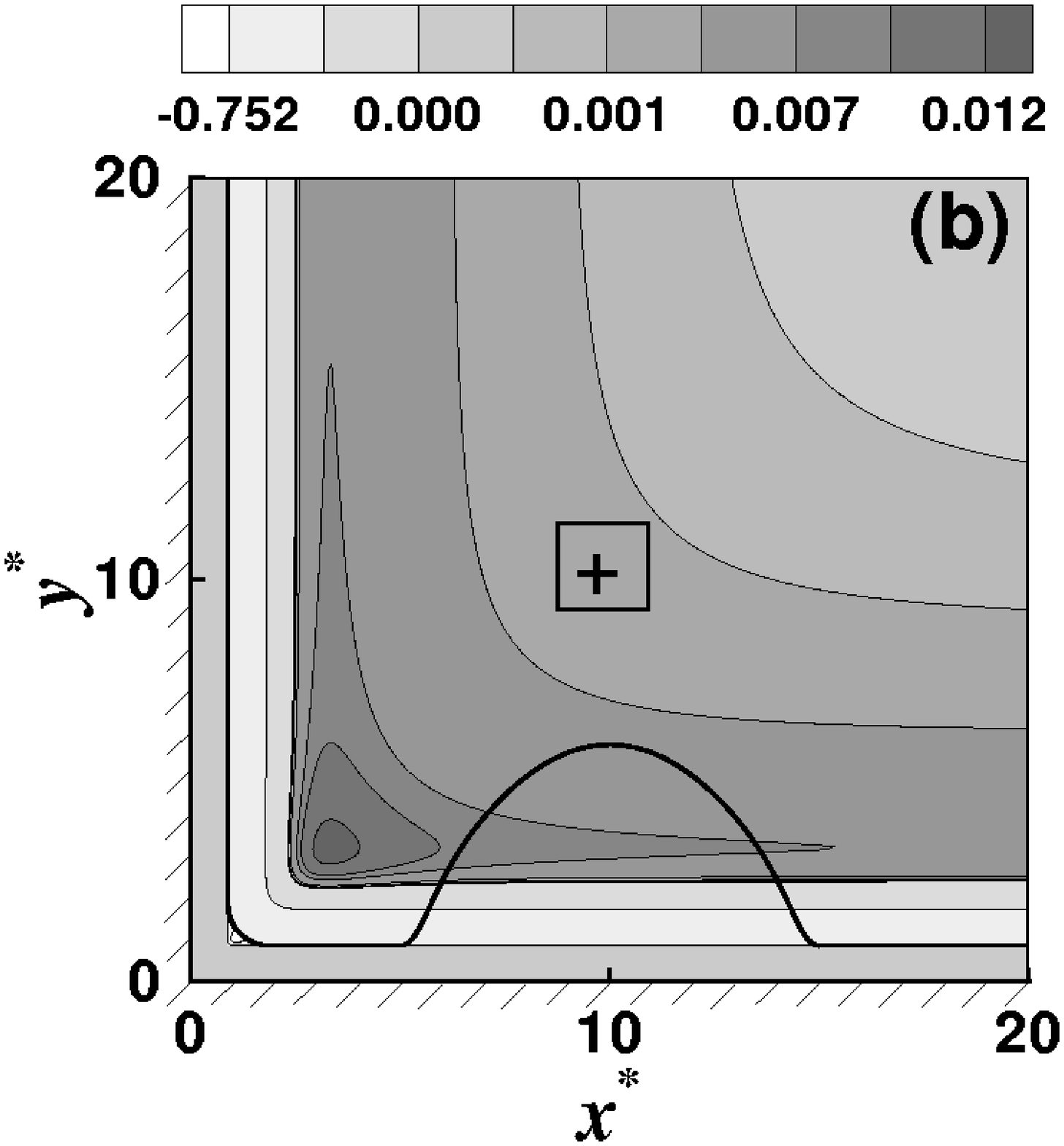}
\caption{\label{fig5} Wedge configuration.
A nanodroplet is placed in front of the wedge and its motion under
the action of the laterally varying DJP is followed. Contour plots of
the DJP of the system for the minus ((a), $B=-1$, $C=1$) and the plus
((b), $B=-2.5$, $C=1$) case are shown.} 
\end{figure}

\begin{figure}
\includegraphics[width=0.49\linewidth]{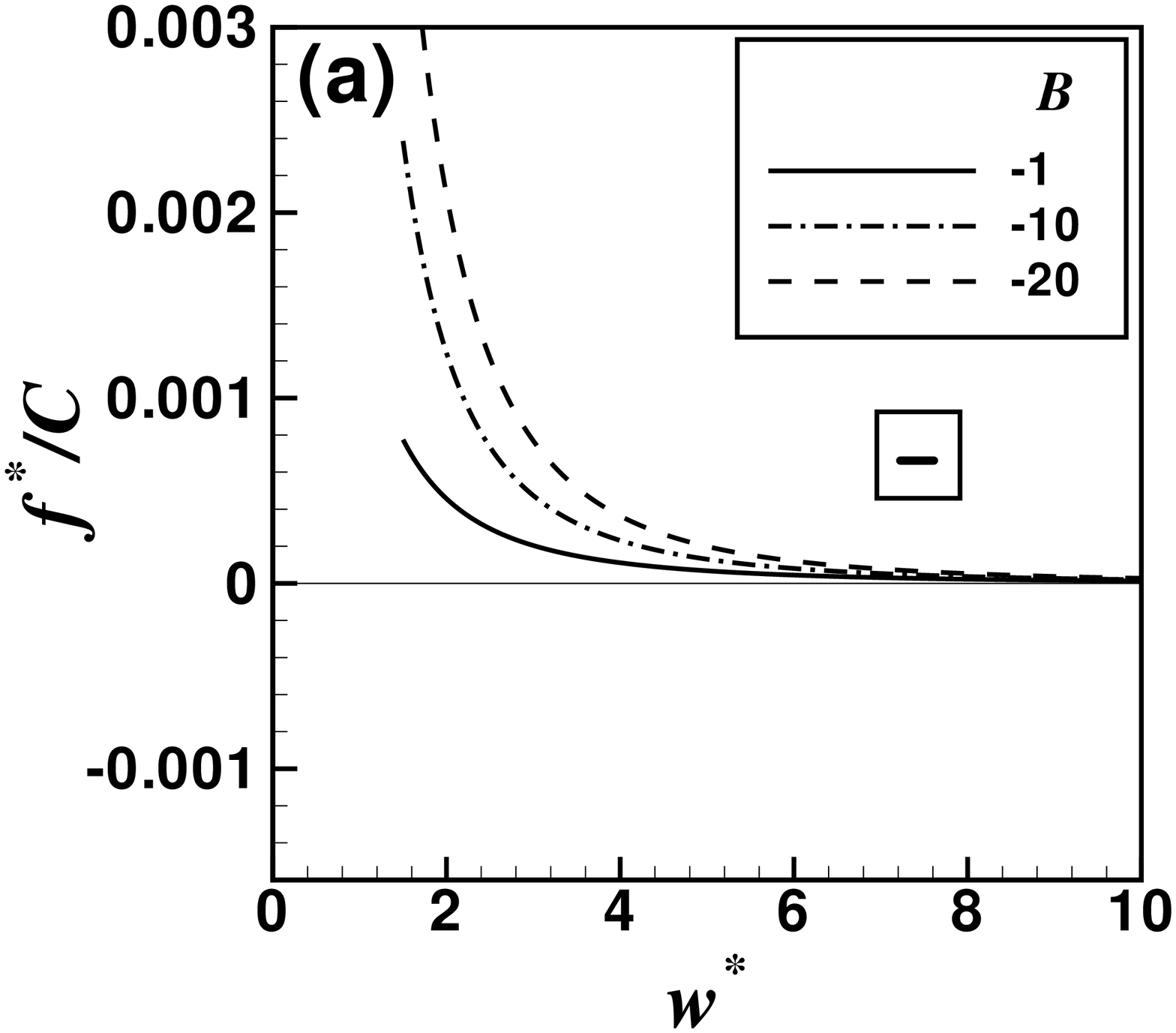}
\includegraphics[width=0.49\linewidth]{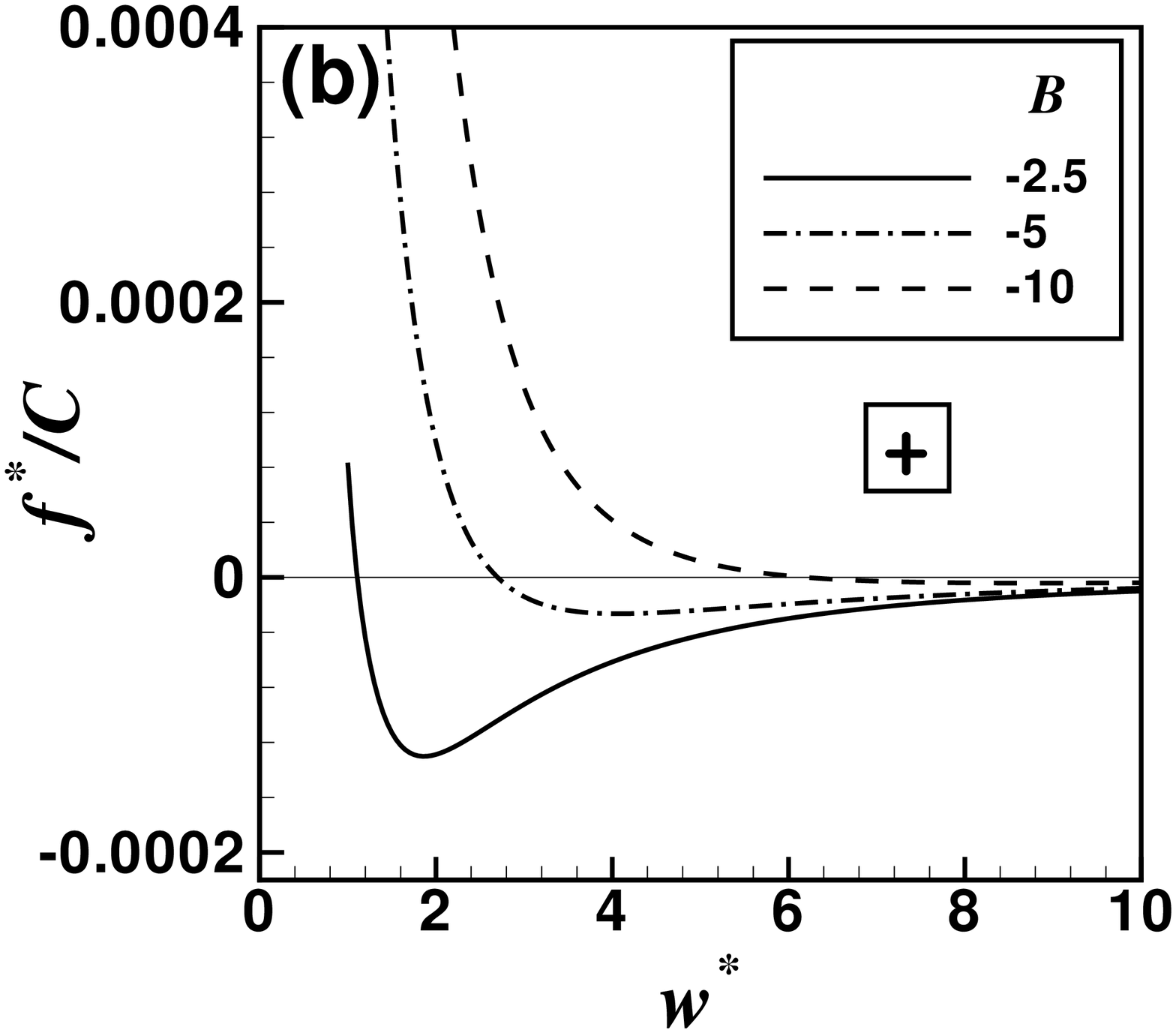}
\caption{\label{fig6}
Normalized DJP induced lateral force densities $f^*$ acting
on an initial droplet configuration of height $a^*=15$ and at distance $w^*$ 
(see Fig.~\ref{fig7}) from a wedge for the minus (a) and the plus (b) case.} 
\end{figure}

\begin{figure}
\includegraphics[width=\linewidth]{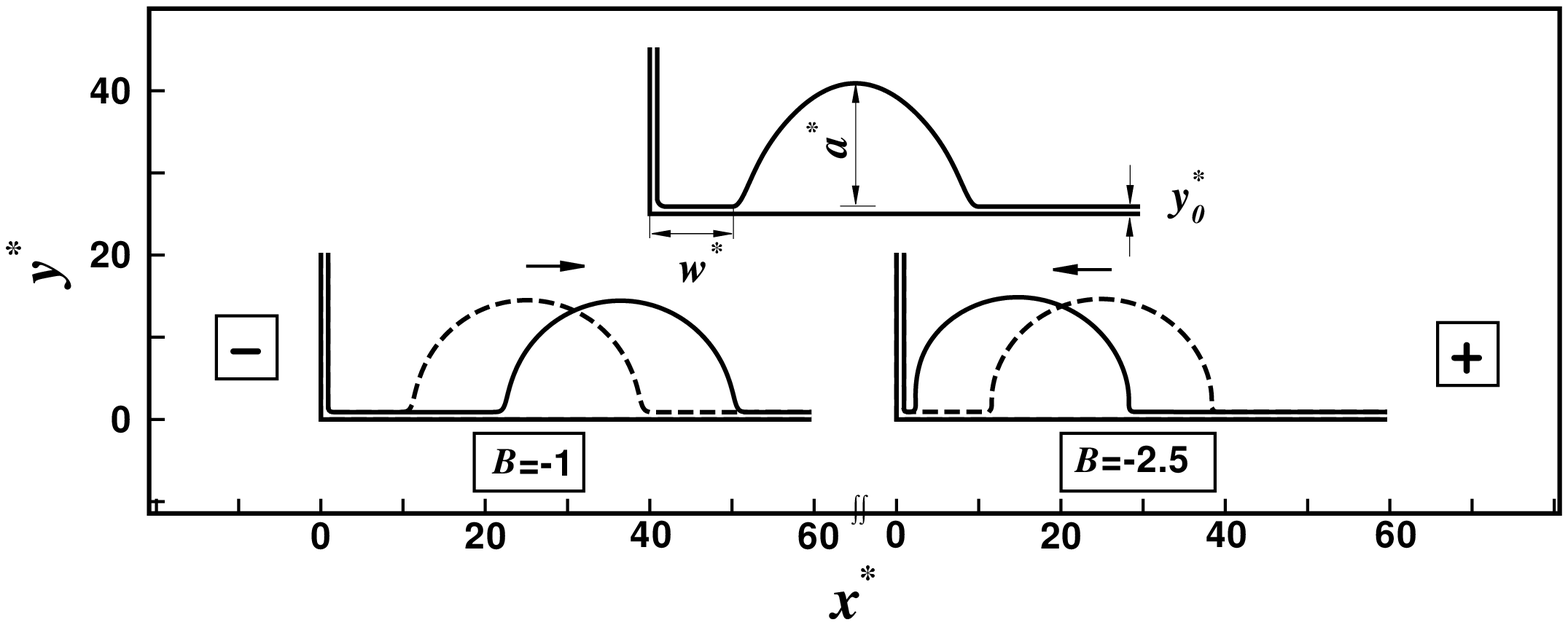}
\caption{\label{fig7} Motion of a droplet ($a^*=15$,
$w^*=10$) near a wedge for the minus (lower
left, $B=-1$, $C=1.27$) and the plus (lower right, $B=-2.5$,
$C=4.23$) case. For $y_0^*$ see the insets in Fig. \ref{fig2}. The upper graph shows the
initial droplet shape. Shown are interface profiles at $t^*=450$ (dashed) and
$t^*=99750$ (solid) for the minus and $t^*=160$ (dashed) and
$t^*=14683$ (solid) for the plus case. The chosen values of $B$ and $C$ provide $\theq=90^\circ$\/.
}
\end{figure}
As a second system we consider a wedge-like substrate.
A wedge can be viewed as being composed of a flat vertical substrate
and an edge. Thus, combining 
their contributions yields
\begin{multline}
\Pi_{cw}^*(x^*,y^*)=\Pi_{cf}^*(x^*)-\Pi_c^{u*}(y^*,x^*)\\
+\Pi_e^*(-x^*,y^*)+\Pi_c^{u*}(-x^*,y^*).
\end{multline}
For a non-coated wedge, within our model this is 
in accordance with Ref. \cite{Napiorkovski:1992}\/.
Typical examples of the DJP in the wedge for the minus and the
plus cases are depicted in Fig.~\ref{fig5}\/. For different values of
$B$ the DJP induced lateral force on a parabolic ridge is shown in 
Figs.~\ref{fig6}(a) and (b) for
the minus and the plus case, respectively. 
The force is always positive, i.e., pointing away from the wedge for the
minus case with its strength decreasing with distance. Thus, one expects a
droplet to move away from the wedge. For the plus case  the force changes
sign from negative to positive near the wedge at a distance which
increases with decreasing $B$\/. Thus one expects 
the droplet to move towards the wedge and then 
to stop before reaching the wedge. This is indeed what is observed in the numerical
calculations of a a liquid ridge as shown in Fig.~\ref{fig7}\/. 
The droplet moves away from the wedge with
decreasing speed in the minus case and towards the wedge with
increasing speed in the plus case. However, in the plus
case the droplet stops before it
reaches the wedge. Only if the droplet is driven into the
wedge (e.g., by external forces) or if it touches the
wedge during the initial relaxation process, the droplet gets
trapped in the wedge and forms a configuration symmetric with respect
to the bisector of the wedge.

In summary we have shown how topographic substrate features generate
motion of non-volatile nanodroplets residing in their vicinity.
This motion depends on the details of the interplay
between the liquid-liquid and the liquid-substrate interactions. 
On substrates for which in the planar case the DJP 
approaches zero from above at large distances (plus
case), droplets move towards edges and away from wedges, and vice versa in
the opposite case (minus case)\/.
This resembles similar phenomena
occurring near chemical heterogenities \cite{Chaudhury:1992}\/. 

Taking 1 nm and 0.02 N/m as typical values of $b$ and $\sigma$, respectively, 
and for the values of $C$ and $B$ chosen in Figs.~\ref{fig4} and \ref{fig7}, our
model calculations predict that near an 
edge the DJP induced lateral force on a liquid
ridge 30~nm long, 15~nm high, and 1~nm thick is of the order of
$10^{-13}$~N\/. For this droplet size the gravitational force due to a
possible tilting of the substrate is about eight orders of magnitude smaller. 
Only for micron-sized drops the gravitational body force becomes competitive. 
The time scale $\mu/A$ for the motion is roughly 
$[10^{-8}\mu/(\textnormal{Pa s)}]$ s so that the average velocities 
for the cases studied are ca. 
$\left\lbrace 10^{-5}/[\mu/(\textnormal{Pa s})]\right\rbrace $ m/s. 
Taking $\mu$ between 0.1 Pa s and 100 Pa s 
(different Polydimethylsiloxanes at ambient temperature) the velocity ranges 
from 0.1 mm/s to 0.1 $\mu$m/s, which is comparable with the velocities of droplets 
of the same kind of liquid but exposed to and driven by a chemical step 
generated by a contact angle contrast of a few degrees \cite{Raphael:1988}. 
Our calculations indicate that the initial relaxation of the droplet shape 
from the prepared one, i.e., spreading of the droplet radius, 
occurs faster than the subsequent motion of the center of gravity of the droplets. 
Since the spreading rate of the monolayer precursor film outside the droplet tends 
to be larger than that of the apparent droplet radius \cite{Heine:2005} 
our calculations lead to the expectation that the velocity for 
the precursor spreading is larger than the motion of the droplets.

Our findings have a bearing on the distribution of droplets after 
condensating liquid onto nano-sculptured substrates and on breath figures, 
in addition to possible implications for the design of open 
nanofluidic systems in which the activating force, 
partially or totally, is provided by the system itself. 

\acknowledgments{M. R. acknowledges financial support by the
Deutsche Forschungsgemeinschaft (DFG) within the priority
program SPP  1164 under grant number RA 1061/2-1.}

\end{document}